\documentclass[aps,prl,twocolumn,showpacs,superscriptaddress]{revtex4}

\usepackage{graphicx}% Include figure files
\usepackage{dcolumn}% Align table columns on decimal point
\usepackage{bm}% bold math
\def\lsim{\lower.5ex\hbox{$\; \buildrel < \over \sim \;$}}
\def\gsim{\lower.5ex\hbox{$\; \buildrel > \over \sim \;$}}
\newcommand{\msun}{M$_\odot$}
\newcommand{\msuny}{M$_\odot$~yr$^{-1}$}

\begin{document}

%\preprint{}

\title{Hydrogen Burning of $^{17}$O in Classical Novae}



\author{A. Chafa}
\affiliation{USTHB-Facult\'e de Physique, BP 32, El-Alia, 16111 Bab Ezzouar,
Algiers, Algeria}
%\thanks{}

\author{V. Tatischeff}
\affiliation{CSNSM, IN2P3-CNRS and Universit\'e Paris-Sud, F-91405 Orsay Cedex,
France}

\author{P. Aguer}
\affiliation{CENBG, IN2P3-CNRS and Universit\'e de Bordeaux I, F-33175
Gradignan, France}

\author{S. Barhoumi}
\affiliation{UMBM, B.P. 166, Route ICHBILLIA, 28000 M'sila, Algeria}

\author{A. Coc}
\affiliation{CSNSM, IN2P3-CNRS and Universit\'e Paris-Sud, F-91405 Orsay 
Cedex, France}

\author{F. Garrido}
\affiliation{CSNSM, IN2P3-CNRS and Universit\'e Paris-Sud, F-91405 Orsay 
Cedex, France}

\author{M. Hernanz}
\affiliation{Institut de Ci\`encies de l'Espai (CSIC), and Institut d'Estudis
Espacials de Catalunya, E-08034 Barcelona, Spain}

\author{J. Jos\'e}
\affiliation{Departament de F\'isica i Enginyeria Nuclear (UPC) and Institut 
d'Estudis Espacials de Catalunya, E-08034 Barcelona, Spain}

\author{J. Kiener}
\affiliation{CSNSM, IN2P3-CNRS and Universit\'e Paris-Sud, F-91405 Orsay 
Cedex, France}

\author{A. Lefebvre-Schuhl}
\affiliation{CSNSM, IN2P3-CNRS and Universit\'e Paris-Sud, F-91405 Orsay 
Cedex, France}

\author{S. Ouichaoui}
\affiliation{USTHB-Facult\'e de Physique, BP 32, El-Alia, 16111 Bab Ezzouar,
Algiers, Algeria}

\author{N. de S\'er\'eville}
\affiliation{CSNSM, IN2P3-CNRS and Universit\'e Paris-Sud, F-91405 Orsay 
Cedex, France}
\affiliation{Universit\'e Catholique de Louvain, Chemin du Cyclotron 2, 
B-1348 Louvain-la-Neuve, Belgium}

\author{J.-P. Thibaud}
\affiliation{CSNSM, IN2P3-CNRS and Universit\'e Paris-Sud, F-91405 Orsay 
Cedex, France}


\date{\today}

\begin{abstract}
We report on the observation of a previously unknown resonance at
$E_R^{lab}$=194.1$\pm$0.6 keV in the $^{17}$O($p$,$\alpha$)$^{14}$N reaction,
with a measured resonance strength $\omega\gamma_{p\alpha}$=1.6$\pm$0.2 meV. 
We studied in the same experiment the $^{17}$O($p$,$\gamma$)$^{18}$F reaction
by an activation method and the resonance-strength ratio was found to be
$\omega\gamma_{p\alpha}$/$\omega\gamma_{p\gamma}$=470$\pm$50. The 
corresponding excitation energy in the $^{18}$F compound nucleus was 
determined to be 5789.8$\pm$0.3 keV by $\gamma$-ray measurements using the 
$^{14}$N($\alpha$,$\gamma$)$^{18}$F reaction. These new resonance properties 
have important consequences for $^{17}$O nucleosynthesis and $\gamma$-ray 
astronomy of classical novae.
\end{abstract}


%\keywords{}

\maketitle

Classical novae are caused by thermonuclear runaways that occur on hydrogen
accreting white dwarfs in close binary systems. They are thought to be a 
major source of the oxygen rarest isotope, $^{17}$O \cite{jos98,sta98} and 
to synthesize the radioisotope $^{18}$F ($T_{1/2}$=110~min), whose 
$\beta^+$-decay produces a $\gamma$-ray emission that could be detected 
with the INTEGRAL observatory or with future $\gamma$-ray satellites 
\cite{her99,her04}. However, both the $^{17}$O and $^{18}$F productions 
strongly depend on the $^{17}$O($p$,$\alpha$)$^{14}$N and 
$^{17}$O($p$,$\gamma$)$^{18}$F thermonuclear rates, whose precise knowledge is
thus required in the range of temperatures attained during nova outbursts 
((1--4)$\times$10$^8$~K).

A new resonance at $E_R^{cm}$$\simeq$180~keV was recently observed in the
$^{17}$O($p$,$\gamma$)$^{18}$F reaction \cite{fox04}. With a measured 
resonance strength of (1.2$\pm$0.2)$\times$10$^{-6}$~eV, the uncertainty of 
the $^{17}$O($p$,$\gamma$)$^{18}$F reaction rate was reduced by orders of
magnitude at nova temperatures. This resonance corresponds to a level of 
spin-parity J$^\pi$=2$^-$ in the $^{18}$F compound nucleus, which was 
previously observed at an excitation energy $E_x$=5786$\pm$2.4~keV via the
$^{14}$N($\alpha$,$\gamma$)$^{18}$F reaction \cite{rol73,til95}. In the 
latter experiment, the lifetime of this state was found to be
$\tau$=15$\pm$10~fs from a measurement based on the Doppler-shift attenuation 
method \cite{rol73}. This measurement led to the somewhat surprising 
ratio $\Gamma_\gamma / \Gamma_\alpha$ $\sim$1, which was used for estimating 
the ($p$,$\alpha$) resonance strength \cite{fox04,ang99}. We
performed a new experimental study of this state using the 
$^{14}$N($\alpha$,$\gamma$)$^{18}$F reaction, in order to specify both its 
excitation energy and width. We then measured in a second experiment
the strengths of the resonances at $E_R^{cm}$$\simeq$180~keV in both the 
$^{17}$O($p$,$\gamma$)$^{18}$F and $^{17}$O($p$,$\alpha$)$^{14}$N reactions. 
For the latter reaction, the resonance was never observed before our 
measurements. 

The first experiment was performed at the 4 MV Van de Graaff accelerator of 
the CENBG laboratory (Bordeaux), with an $^4$He beam of energy
$E_\alpha$=1775~keV. Typical beam intensities on target were 20--30 $\mu$A with 
a 3-mm$\times$3-mm spot. We used three TiN targets, which were fabricated by 
nitration in an atmosphere of purified N of a Ti layer evaporated on a thick 
Cu backing. The targets were thick enough ($\gsim$250 $\mu$g/cm$^2$) to allow 
the simultaneous excitation of four levels in $^{18}$F between 5.6 and 5.8~MeV. 
The $\gamma$-rays were detected with three large volume, high purity Ge 
detectors placed horizontally at $\sim$9~cm from the target, at the 
laboratory angles $\theta_{lab}$=0$^\circ$, 123$^\circ$ and 144$^\circ$. The 
Ge detector at 0$^\circ$ was actively shielded with BGO scintillation detectors 
for Compton suppression. Three radioactive sources of $^{137}$Cs, $^{60}$Co and $^{88}$Y 
were permanently placed near the target to measure standard calibration lines 
together with the beam-induced $\gamma$-rays. Taking advantage of boron 
contamination in the target, we also included in the set of calibration lines 
a $^{13}$C, Doppler-unshifted line at 3853.170$\pm$0.022~keV, produced by the 
$^{10}$B($\alpha$,$p$)$^{13}$C reaction. 

A measured $\gamma$-ray spectrum is shown in Fig.~1. The two lines from the 
decay of the $E_x$=5789.8 keV state to the lower-lying levels at $E_x$=937.2 
and 1080.54~keV are clearly observed. The excitation energy and lifetime of 
the level of interest were determined from the measurements at the three 
detection angles of the energy differences between these two lines and 
adjacent lines arising from the decay of the lower resonant state in 
$^{18}$F at $E_x$=5671.6$\pm$0.2~keV \cite{bog89}. Because the lifetime of 
the 5671.6 keV level is very short \cite{til95}, its deexcitation 
$\gamma$-rays are affected by a full Doppler shift. The attenuation factor 
for the Doppler shift of the $\gamma$-rays produced by the decay of the 
5789.8 keV state was found to be $f_{a}$$>$0.9925 (1$\sigma$-limit). The 
corresponding mean lifetime of the decaying level was obtained from detailed 
Monte-Carlo simulations for the slowing down of the recoiling excited 
$^{18}$F nuclei in the target material and the subsequent $\gamma$-ray 
emission and detection with our experimental setup (see as an example the 
inset of Fig.~1). The result is $\tau$$<$2.6~fs, in disagreement with 
Ref.~\cite{rol73}. The excitation energy was found to be 
$E_x$=5789.8$\pm$0.3~keV, in agreement with the very recent result of 
Ref~\cite{fox05}. Further details on the data analysis and the Monte-Carlo 
simulations will be given in a forthcoming publication \cite{cha05}. 

\begin{figure}
\includegraphics[width=7.5cm]{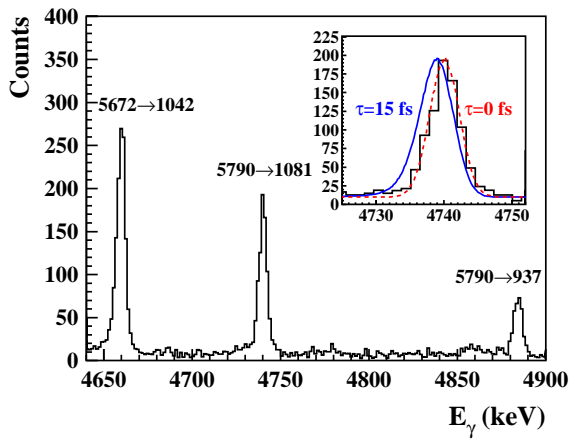}%
\caption{Relevant part of a sample $\gamma$-ray spectrum obtained in the 
$^{14}$N($\alpha$,$\gamma$)$^{18}$F experiment with the Ge detector at 
$\theta_{lab}$=0$^\circ$. The inset shows a comparison of the observed 
full-energy peak for the 5789.8$\rightarrow$1080.54 keV transition in 
$^{18}$F with simulated line emissions calculated for a mean lifetime $\tau$ 
of the 5789.8 keV state equal to 15~fs (solid curve) and 0~fs (dashed curve).
\label{fig1}}
\end{figure}

The search for the previously unknown resonance in the 
$^{17}$O($p$,$\alpha$)$^{14}$N reaction at $E_R^{lab}$=194.1$\pm$0.6~keV 
\cite{aud03} was motivated by the relatively large width found for the 
corresponding resonant state at $E_x$=5789.8~keV in $^{18}$F: 
$\Gamma$$>$250~meV. This second experiment was carried out at 
the electrostatic accelerator PAPAP of the CSNSM laboratory (Orsay), which 
supplies intense proton beams of energies $E_p$$<$250~keV. Beam 
currents of 60--90~$\mu$A were typically sustained on the target, which was 
cooled with deionized water. An annular electrode biased at a voltage of 
-300~V was mounted in front of the target to suppress the escape of the 
secondary electrons. The target holder was surrounded with a copper 
assembly, cooled with liquid nitrogen to limit the carbon buildup 
during the irradiation. 

The outgoing $\alpha$-particles were detected with 4 passivated implanted 
planar silicon detectors with active areas of 3~cm$^2$. They were 
placed at a distance of 14~cm from the target and at the laboratory angles 
$\theta_{lab}$=105$^\circ$, 120$^\circ$, 135$^\circ$ and 150$^\circ$. 
Each detector was shielded by a 2~$\mu$m thick foil of aluminized Mylar in 
order to stop the intense flux of elastically scattered protons. The detector 
solid angle was measured with two calibrated $\alpha$-sources: $^{241}$Am and 
a mixing of $^{239}$Pu, $^{241}$Am and $^{244}$Cm. The average solid angle per 
detector, $\Omega_{lab}$, was found to be (1.42$\pm$0.04)$\times$10$^{-2}$ sr. 
A sample spectrum is shown in Fig.~2a.

The strength of the $^{17}$O($p$,$\alpha$)$^{14}$N resonance was determined
relative to that of the well-known resonance at $E_R^{lab}$=150.9~keV in the 
$^{18}$O($p$,$\alpha$)$^{15}$N reaction. Targets enriched in $^{17}$O or 
$^{18}$O were produced by ion implantation in 0.3~mm thick Ta sheets at the 
SIDONIE implanter of the CSNSM, using the same experimental procedure. The 
total irradiation fluence was 1.5$\times$10$^{18}$ atoms~cm$^{-2}$, equally 
distributed at 30, 10 and 2.5~keV implantation energies. The targets were 
analyzed by Rutherford backscattering spectrometry (RBS) measurements 
performed at the ARAMIS accelerator (CSNSM) with a $^4$He beam of 1.2~MeV 
energy. No difference could be observed between the $^{17}$O- and 
$^{18}$O-implanted targets. In particular, a similar stoichiometry was found 
for the $^{17}$O and $^{18}$O targets whatever the depth and surface position, 
with a maximum ratio (O/Ta)$_{\rm max}$=3.1$\pm$0.3. No change in 
the target stoichiometry could be noted from RBS measurements performed after 
the proton irradiation, where the charge accumulated on each target was 
typically 1~C. 

The similarity of the $^{17}$O- and $^{18}$O-implanted targets can be seen 
in Fig.~2b, which shows a comparison of yield data obtained for 
the new resonance at $E_R^{lab}$=194.1~keV in the 
$^{17}$O($p$,$\alpha$)$^{14}$N reaction and for the 150.9~keV resonance 
of $^{18}$O($p$,$\alpha$)$^{15}$N. The observed depth profile is well 
explained by the implantation procedure. The yield measurements were repeated 
with three $^{17}$O- and two $^{18}$O-implanted targets and fully compatible 
results were obtained. 

\begin{figure}
\includegraphics[width=6.cm]{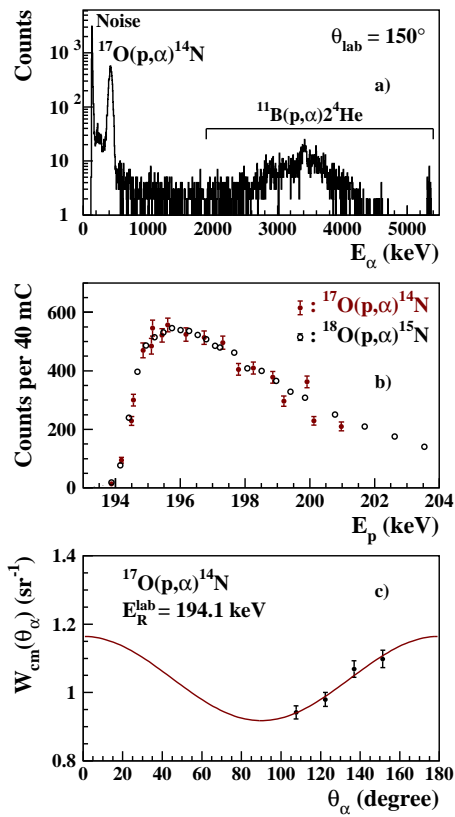}%
\caption{(a) Sample $\alpha$ spectrum obtained in the 
$^{17}$O($p$,$\alpha$)$^{14}$N experiment, for $E_p$=196.5~keV and an 
accumulated charge of 0.93~C. (b) Yield data for the new resonance at 
$E_R^{lab}$=194.1~keV in the $^{17}$O($p$,$\alpha$)$^{14}$N 
reaction (filled symbols) and the well-known $^{18}$O($p$,$\alpha$)$^{15}$N 
resonance at $E_R^{lab}$=150.9~keV (empty symbols). The data for this latter 
resonance were appropriately normalized and shifted in energy to be compared 
with those obtained with the $^{17}$O target. (c) $\alpha$ angular 
distribution (center-of-mass) for the $^{17}$O($p$,$\alpha$)$^{14}$N 
resonance. The solid line shows a Legendre-polynomial fit to the data, 
$W_{cm}(\theta_{\alpha})=1+a_2P_2({\rm cos}\theta_{\alpha})$, which yields 
$a_2$=0.16$\pm$0.03.
\label{fig2}}
\end{figure}

The strength of the $^{17}$O($p$,$\alpha$)$^{14}$N resonance was deduced from 
the relation
\begin{equation}
\omega\gamma_{p\alpha}=\omega\gamma_{p\alpha}^{18}
{M_{17} \over M_{17}+m_p}{M_{18}+m_p \over M_{18}}{E_R^{17} \over E_R^{18}}
{\epsilon_{17} \over \epsilon_{18}}{Y_{p\alpha}^{17} \over Y_{p\alpha}^{18}}~.
\end{equation}
Here, $\omega\gamma_{p\alpha}^{18}$=0.167$\pm$0.012 eV is the strength of the 
$^{18}$O($p$,$\alpha$)$^{15}$N resonance at $E_R^{lab}$=150.9~keV 
\cite{lor79,bec95}; $m_p$, $M_{17}$ and $M_{18}$ are the proton, $^{17}$O and 
$^{18}$O masses, respectively; $E_R^{17}$ and $E_R^{18}$ are the laboratory 
energies of the two resonances; 
$(\epsilon_{17} / \epsilon_{18})$=0.95$\pm$0.05 is the ratio of the effective 
stopping-powers \cite{zie}, where the error arises from the uncertainty in 
the stoichiometry of the $^{17}$O and $^{18}$O targets; and 
$(Y_{p\alpha}^{17} / Y_{p\alpha}^{18})$=(7.7$\pm$0.9)$\times$10$^{-3}$ 
is the ratio of the measured reaction yields for the two resonances, where 
the main uncertainties come from the target composition (10\%) and the beam 
current integration (5\%). For the determination of $Y_{p\alpha}^{17}$, we 
took into account the measured $\alpha$-particle angular distribution shown in 
Fig.~2c. Equation~1 gives $\omega\gamma_{p\alpha}$=1.6$\pm$0.2~meV.

\begin{figure}[b!]
\includegraphics[width=6.2cm]{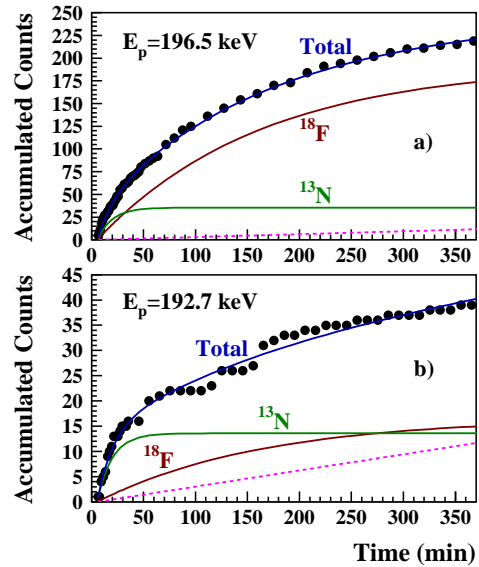}%
\caption{Measured $\beta^+$ activities of two $^{17}$O-implanted targets 
irradiated at (a) $E_p$=196.5~keV (on-resonance) and (b) $E_p$=192.7~keV 
(off-resonance). The time origin corresponds to the stopping of the proton 
irradiation. Also shown are the fitted $^{18}$F and $^{13}$N decay 
curves (solid lines) and the measured background radiation (dashed lines). 
\label{fig3}}
\end{figure}

The strength of the $^{17}$O($p$,$\gamma$)$^{18}$F resonance at 
$E_R^{lab}$=194.1~keV was obtained from measurements by an activation 
method of the $^{18}$F total production in irradiated $^{17}$O targets. 
The $^{18}$F $\beta^+$-activity was measured with two large volume Ge 
detectors positioned opposite to one another in a very close geometry, in 
order to register in time coincidence the two 511~keV photons from 
positron-electron annihilation. The efficiency for the $\beta^+$-activity 
detection was measured with a calibrated $^{22}$Na source to be 
$\epsilon_{\beta^+}$=(2.7$\pm$0.1)\%. The $^{17}$O-implanted targets were 
bombarded for $\sim$5 hours at $\sim$70~$\mu$A beam intensity and then 
rapidly placed between the two Ge detectors. $\alpha$-particle yields 
were continuously recorded during the irradiation. 

Apart from $^{18}$F ($T_{1/2}$=109.77~min), only two long-lived positron 
emitters could be significantly produced during the irradiation phase: 
$^{11}$C ($T_{1/2}$=20.39~min) by the reaction $^{10}$B($p$,$\gamma$)$^{11}$C 
and $^{13}$N ($T_{1/2}$=9.965~min) by the reaction 
$^{12}$C($p$,$\gamma$)$^{13}$N. The former reaction was found to be 
negligible, from systematic measurements of the boron contamination in the 
targets via the $^{11}$B($p$,$\alpha$)2$^{4}$He reaction (see Fig.~2a). But a 
relatively small production of $^{13}$N had to be taken into account, because 
of a carbon buildup on target of about 0.5~$\mu$g~cm$^{-2}$ per Coulomb of 
accumulated proton charge. A blank Ta target irradiated in the same 
experimental conditions showed no activity but the one of $^{13}$N.

We used the $^{12}$C($p$,$\gamma$)$^{13}$N reaction to test the experimental 
setup. A C target of 20~$\mu$g~cm$^{-2}$ evaporated on a Ta sheet was 
irradiated for 30~min at $E_p$=196~keV. The astrophysical S-factor derived 
from the measured $^{13}$N activity is 4.0$\pm$0.8~keV~b, in good agreement 
with previous results \cite{rol74}.

Figure~3 compares the measured activities of two $^{17}$O targets irradiated 
at $E_p$=196.5 and 192.7~keV. The fitted $^{18}$F and $^{13}$N decay curves 
were obtained from the maximum of the likelihood function for 
Poisson-distributed data. The total numbers of $^{18}$F nuclei contained in 
the targets at the end of the proton irradiation were 7160$\pm$700 for 
$E_p$=196.5~keV (with $T_{1/2}$=105$^{+19}_{-14}$~min for the fitted $^{18}$F
half-life) and 610$\pm$260 for $E_p$=192.7~keV 
(with $T_{1/2}$=103$^{+217}_{-47}$~min). The larger number of $^{18}$F nuclei 
produced at the highest beam energy is clearly due to the excitation of the 
$^{17}$O($p$,$\gamma$)$^{18}$F resonance at $E_R^{lab}$=194.1~keV. The 
$^{18}$F production at $E_p$=192.7~keV results from the direct capture (DC)
process interfering with the low-energy tail of the studied resonance. Our
measurement agrees within large statistical uncertainties with the DC 
evaluation of Ref.~\cite{rol75}. To derive the resonance strength, a small 
contribution of (6$\pm$3)\% for the DC process was subtracted from the 
$^{18}$F total production at $E_p$=196.5~keV.

\begin{figure}
\includegraphics[width=6.5cm]{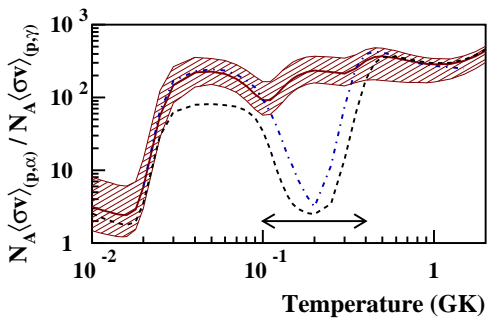}%
\caption{Ratio of the $^{17}$O($p$,$\alpha$)$^{14}$N and 
$^{17}$O($p$,$\gamma$)$^{18}$F reaction rates (solid line with hatched area 
reflecting uncertainties), in comparison with the previous results of Angulo 
{\it et al.} \cite{ang99} (dashed line) and Fox {\it et al.} \cite{fox04} 
(dotted-dashed line). The horizontal arrow shows the range of typical novae 
temperatures.
\label{fig4}}
\end{figure}

We also applied to the measured ($p$,$\gamma$) resonance strength a 
correction arising from the backscattering of $^{18}$F nuclei, which could 
escape the target at the time of their production. This correction was 
evaluated to be (4$\pm$2)\% from calculations performed with the program 
SRIM-2003 \cite{zie} for the observed implantation profile of the $^{17}$O 
targets. We finally obtained from the weighted mean of two compatible 
measurements at $E_p$=196.5~keV: 
$\omega\gamma_{p\alpha}$/$\omega\gamma_{p\gamma}$=470$\pm$50. Here, the error 
mainly arises from uncertainties in the measured $\beta^+$ activities (7\%) 
and the associated detection efficiency (5\%), as well as in the measured 
$\alpha$-particle intensities (1\%) and detection efficiency (3\%). The 
resulting ($p$,$\gamma$) resonance strength is 
$\omega\gamma_{p\gamma}$=(3.4$\pm$0.6)$\times$10$^{-6}$~eV, which is 
significantly larger than the value of Ref.~\cite{fox04}: 
$\omega\gamma_{p\gamma}$=(1.2$\pm$0.2)$\times$10$^{-6}$~eV (see 
Ref.~\cite{cha05} for a discussion of the discrepancy). 

We calculated the thermonuclear rates of the $^{17}$O+$p$ reactions by using 
our experimental results. These calculations will be discussed in detail in 
Ref.~\cite{cha05}. We show in Fig.~4 the ratio of the ($p$,$\alpha$) and 
($p$,$\gamma$) reaction rates. The comparison with previous estimates 
\cite{ang99,fox04} essentially illustrates the strong effect of the 
previously unknown resonance in the $^{17}$O($p$,$\alpha$)$^{14}$N reaction, 
for stellar temperatures of $\sim$(1--4)$\times$10$^8$~K. With the present
results, the $^{17}$O+$p$ rates are now established from measured nuclear data
in the whole range of nova temperatures.

The impact of the new $^{17}$O+$p$ rates has been studied by means of three 
hydrodynamic simulations of classical nova outbursts. For illustrative 
purposes, we have adopted a model of 1.15 \msun\ ONe white dwarf accreting 
hydrogen-rich material at a rate of $2 \times 10^{-10}$ \msuny, with three 
different choices for the $^{17}$O+$p$ rates: from Refs.~\cite{ang99,fox04} 
and this work. With the new $^{17}$O+$p$ rates, we found the final $^{17}$O 
abundance to be reduced by a factor of $2.4$ with respect to Ref.~\cite{fox04} 
(or 1.4 with Ref.~\cite{ang99} rates). The dramatic increase in 
the new ($p$,$\alpha$) rate reduces the final abundance of $^{18}$F 
by a factor of 2.9 with respect to Ref.~\cite{fox04} (or 7.9 with 
Ref.~\cite{ang99} rates). This translates into a significant reduction of the 
detectability distances (assuming that the flux scales with the $^{18}$F 
yield) by a factor of $\sim$1.7 with respect to Ref.~\cite{fox04} (or 2.8 
with Ref.~\cite{ang99} rates). These new results have also a significant 
impact in other astrophysical topics (although a deeper study with a larger 
number of models is required to properly address this issue): this includes 
the extent of the nova contribution to the Galactic $^{17}$O and estimates 
of oxygen isotopic ratios in the nova ejecta to derive the expected 
composition of presolar oxide grains. 

We wish to thank E. Virassamynaiken and D. Ledu for the preparation of the 
TiN and O-implanted targets, respectively, as well as the technical staff of 
the CSNSM for the constant help. This work was supported by the CMEP 
agreement between France and Algeria under the project No. 01MDU518 and by
AYA2004-06290-C02.


\begin{thebibliography}{01}
\bibitem{bog89} We corrected in this experiment the excitation energy of this 
level, which was previously found at $E_x$=5672.57$\pm$0.32~keV by G. 
Bogaert {\it et al.}, Phys. Rev. C {\bf 39}, 265 (1989). 
The discrepancy is 
probably due to the erroneous value adopted by these authors for the energy 
of the calibration line at 3853.170$\pm$0.022 keV from the 
$^{10}$B($\alpha$,$p$)$^{13}$C reaction.
\end{thebibliography}

\end{document}